\documentclass[qsecnum,amsmath,preprintnumbers,superscriptaddress,nofootinbib,aps,prd,10pt,a4paper]{revtex4-1}
\pdfoutput=1
\usepackage[utf8]{inputenc}
\usepackage{amsmath}
\usepackage{amsfonts}
\usepackage{amssymb}
\usepackage{amsthm}
\usepackage{hyperref}
\usepackage{xcolor}
\usepackage{bm}
\usepackage{graphicx}
\usepackage{feynmp-auto}
\usepackage{mathrsfs}
\usepackage{caption}
\usepackage{subcaption}
\usepackage{framed}

 \def\be{\begin{equation}}
\def\ee{\end{equation}}
 \def\ba{\begin{align}}
\def\ea{\end{align}}
\def\bea{\begin{eqnarray}}
\def\eea{\end{eqnarray}}

\newcommand{\bseq}{\begin{subequations}}
\newcommand{\eseq}{\end{subequations}}

\begin{document}
\preprint{Imperial/TP/2022/AAT/1}
\title{{\bf UV graviton scattering and positivity bounds from IR dispersion relations}}
\author{M. Herrero-Valea}
\email[]{mherrero@ifae.es}
\address{Institut de Fisica d’Altes Energies (IFAE), The Barcelona Institute of Science and Technology, Campus UAB, 08193 Bellaterra (Barcelona), Spain}

\author{A. S. Koshelev}
\email{ak@inpcs.net}
\affiliation{
Departamento de F\'isica, Centro de Matem\'atica e Aplica\c{c}\~oes (CMA-UBI),\\
Universidade da Beira Interior, 6200 Covilh\~a, Portugal
}

\author{A. Tokareva}
\email{a.tokareva@imperial.ac.uk}

\affiliation{Theoretical Physics, Blackett Laboratory, Imperial College London, SW7 2AZ London, U.K.}

\begin{abstract}
Scattering amplitudes mediated by graviton exchange display IR singularities in the forward limit. This obstructs standard application of positivity bounds based on twice subtracted dispersion relations. Such divergences can be cancelled only if the UV limit of the scattering amplitude behaves in a specific way, which implies a very non-trivial connection between the UV and IR behaviors of the amplitude. We show that this relation can be expressed in terms of an integral transform, obtaining analytic results when $t \log{s}\rightarrow 0$. Carefully applying this limit to dispersion relations, we find that infinite arc integrals, which are usually taken to vanish, can give a non-trivial contribution in the presence of gravity, unlike in the case of finite negative $t$. This implies that gravitational positivity bounds cannot be trusted unless the size of this contribution is estimated in some way, which implies assumptions on the UV completion of gravitational interactions. We discuss the relevance of these findings in the particular case of QED coupled to gravity. 
\end{abstract}

\maketitle

\newpage

\section{Introduction}

The analytic properties of the $S$-matrix are a central element of our understanding of Quantum Field Theory (QFT). Stemming from seminal works on partial wave unitarity in the 1970s \cite{Yndurain:1972ix,Pham:1985cr,Ananthanarayan:1994hf}, there has recently been a modern resurgence of interest in this topic, in connection to the study of Effective Field Theories (EFT) \cite{Adams:2006sv}. $S$-matrix properties can be used to formulate \emph{positivity bounds} for 2-to-2 scattering amplitudes within the physical region for scattered momenta in the forward limit. These are written in terms of dispersion relations which relate the scattering amplitude at a given kinematical point with the integral of its imaginary part along the whole physical region, which is strictly positive from the optical theorem.

Standard applications then follow a top-down reasoning. One first promotes a given EFT to be the low energy expansion of an unknown ultraviolet (UV) complete theory satisfying the usual axioms of unitarity, locality, and Lorentz invariance. Positivity bounds are then valid and applicable to the UV complete theory. However, they can also be evaluated at small center of mass energy, in the infrared (IR) region. There, scattering amplitudes are well approximated by those computed in the EFT. As a consequence, positivity implies constraints on the Wilson coefficients accompanying those relevant operators that contribute to the $S$-matrix elements. This UV-IR connection has been thoroughly used in the literature to constrain, or assess the validity, of many different EFTs, see for example \cite{Cheung:2016wjt,Bonifacio:2016wcb,deRham:2017imi,deRham:2017xox,deRham:2018qqo,Afkhami-Jeddi:2018own,Zhang:2018shp,Bellazzini:2019bzh,Melville:2019wyy,Alberte:2019xfh,Alberte:2019zhd,Kim:2019wjo,Herrero-Valea:2019hde,Remmen:2019cyz,Remmen:2020uze,Wang:2020jxr,deRham:2021fpu,Traykova:2021hbr,Davighi:2021osh,Bern:2021ppb}. 

This approach however, relies on the existence of a mass gap in the spectrum of the EFT, a property needed for the forward limit to be regular. In gapless theories, exchange of massless particles leads to forward limit divergences in the scattering amplitudes, which obstruct a direct application of positivity bounds. This divergence can be relaxed for both scalar and vector degrees of freedom by a proper regularization, but the fundamental problem remains for graviton exchange, which requires an alternative approach. An elegant way out of this is to isolate the divergence in the right hand side of the dispersion relation --which is exact --, so that it can be cancelled against the one in the left hand side \cite{Tokuda:2020mlf}. By doing this, one is left with an approximate positivity bound, which can be mildly violated by terms that become important only at very high energies. Nevertheless, these approximate bounds are still very powerful in constraining many IR proposals for gravitational physics \cite{Aoki:2021ckh,Noumi:2021uuv,Herrero-Valea:2021dry}. This result can also obtained in a different way, based on the impact parameter formulation \cite{Caron-Huot:2021rmr}.

In order for this cancellation to be possible, an assumption about the high energy behavior of graviton scattering has to be done, though. In \cite{Tokuda:2020mlf} this is assumed to be of the Regge form, inspired by the Veneziano formula of string scattering \cite{Veneziano:1968yb}; but we could of course wonder if this result is unique, or if there are other possible UV behaviors that work. This has been partially answered in \cite{Herrero-Valea:2020wxz}, where they show that subtraction of the tree-level divergence requires a linear Regee trajectory at leading order in the high energy region, but nothing has been established so far about sub-leading corrections. This is an important question because, for example, string scattering is not an exact linear Regge trajectory. Sub-leading corrections are always present. Thus, it is natural to ask ourselves: Are these unique? What kind of sub-leading terms allow for well-posed dispersion relations and positivity bounds? Is the scattering of strings the only possible UV behavior of graviton scattering that satisfies this condition? And moreover, are positivity bounds insensitive to this choice? In general, we ask ourselves what is the minimal piece of information about the UV behavior of gravitation needed for dispersion relations to be well-posed.

In this work we try to answer this question by reversing the usual direction of thought in the literature in positivity bounds. By looking at the structure of graviton exchange in the IR limit, and exploiting mathematical properties of the dispersion relation, we constrain the UV behavior of graviton-mediated scattering amplitudes. We arrive to an integral formula that relates the IR structure of forward limit divergences to properties of the UV completion, which is further constrained in the limit $t \log s\rightarrow 0$. We also show how this knowledge modifies the standard derivation of positivity bounds, leading even to undeterminate bounds unless extra assumptions about the UV completion are made.

A recent development in a similar direction -- reverse bootstrapping UV amplitudes from IR properties -- was presented in \cite{Alberte:2021dnj}, where they consider the scattering of photons and gravitons in QED coupled to the Einstein-Hilbert action. Based on one-loop computations done in \cite{Alberte:2020bdz}, they show that parametrically large negative terms appear in the dispersion relation, naively contradicting positivity bounds unless new physics is introduced at relatively low energies. Instead, the authors of \cite{Alberte:2021dnj} argue that the presence of these negative large pieces can, and most likely should, have an origin on non-trivial properties of the UV completion, which might in principle be affected by the presence of light particles such as the electron \cite{Alberte:2020jsk}. In this work we show explicitly how this way of thinking, together with our results about the shape of the UV scattering amplitude, can lead to a resolution of the mentioned tension.

This paper is organized as follows. First, we introduce dispersion relations for theories with graviton exchange in \ref{sec:dispersion_rel}, and we show how cancellation of IR divergences determines several properties of scattering amplitudes in the UV in section \ref{sec:asympt_exp}. Later, in section \ref{sec:arc_int} we show how our findings imply a non-vanishing value for the arc integrals contributing to the dispersion relation, and we discuss how this can solve the conundrum that emerges when applying positivity bounds to gravitational coupled QED in section \ref{sec:QED}. Section \ref{sec:fate_bounds} is devoted to show how positivity bounds might be rendered useless by our results in the presence of gravitation, while in section \ref{sec:strings} we explore an explicit example of a UV completion in the form of string amplitudes, finding agreement with our result. Finally, we draw our conclusions in section \ref{sec:conclusions}.

\section{Dispersions relations with graviton exchange} \label{sec:dispersion_rel}

\begin{figure}
    \centering
    \includegraphics[width=0.4\textwidth]{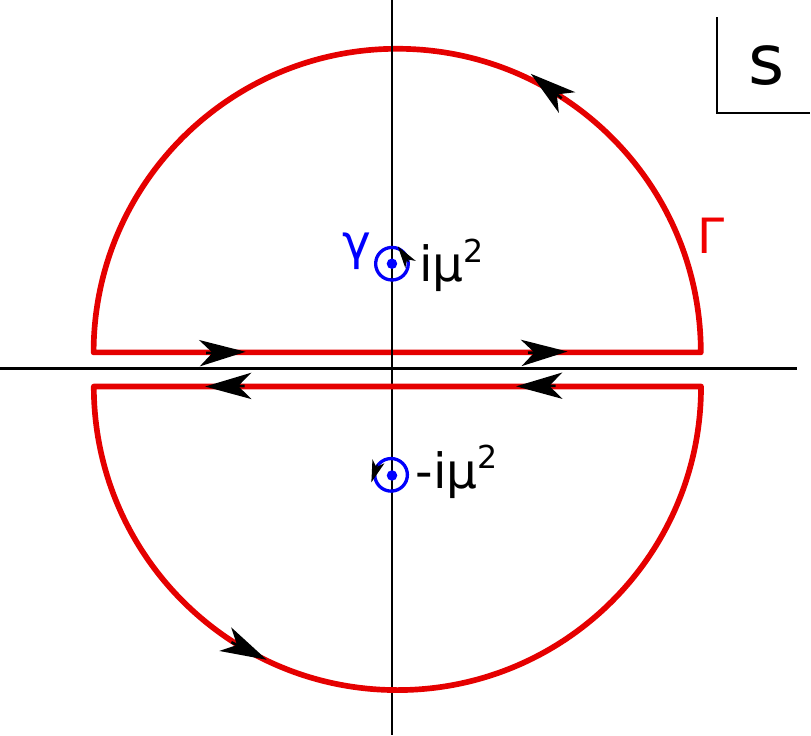}
    \caption{Integration contours for the dispersion relation in \eqref{S} and \eqref{eq:final_disp}.}
    \label{fig:gamma_cont}
\end{figure}

Let us start by considering the $ab\rightarrow ab$ scattering between some identical but otherwise unspecified initial and final states with equal mass $m$, as described in an EFT with an unknown UV completion, but which we demand to be causal, local, unitary and Lorentz invariant. We will also assume that this process includes exchange of a massless graviton. Due to Lorentz invariance, the scattering amplitude ${\cal A}(s,t)$ can be uniquely described in terms of the Mandelstam variables $s$, $t$ and $u$, satisfying $s+t+u=4m^2$. The presence of massless gravitons in the scattering channel implies that the amplitude will diverge in the forward limit\footnote{It is important to remark here that the forward limit corresponds to taking $t\rightarrow 0$ from the negative side of the real axis, since the physical region corresponds to $t<0$.} $t\rightarrow 0^-$. The typical expansion of the amplitude in this limit, including tree-level graviton exchange and one graviton loop, has the form
\begin{equation}
\label{amplitude}
   {\cal A}(s,t)= A_0 \frac{s^2}{M_P^2}\frac{1}{t}+A_1 \frac{s^2}{M_P^4} \log{\left(\frac{-t}{\mu_R^2}\right)}+(\text{regular terms})+(\text{higher loops}),
\end{equation}
where $\mu_R$ is the renormalization scale. In the forward limit, this expression has explicit $1/t$ and $\log{t}$ divergences. The pole is inherited from the one in the graviton propagator, while loop corrections are responsible for generating the logarithmic branch cut, which represents production of soft gravitons of arbitrarily low energy. The values of $A_0$ and $A_1$ characterize the particular theory from which this amplitude is obtained. In perturbation theory, they are proportional to the residue in the pole of the propagator of the massless graviton, and to the $\beta$-function of the $a^2b^2$ coupling. Higher loops -- two and beyond -- will produce further logarithmic divergences, but we ignore them hereinafter, since they are further suppressed. From crossing symmetry, the amplitude contains in general equivalent divergences $s^{-1}$, $u^{-1}$, $\log(s)$, $\log(u)$. Due to the latter, the forward limit amplitude ${\cal A}(s,t\rightarrow 0^-)$ will exhibit also a branch cut along the whole real axis in the complex plane for $s$, which obstructs the standard derivation of positivity bounds \cite{Adams:2006sv}. This can be avoided, however, by following the derivation in \cite{Herrero-Valea:2020wxz}, which we adopt hereinafter. We thus consider the following quantity
\begin{equation}
\label{S}
\Sigma(\mu,0^-)=\frac{1}{2\pi i}\oint_{\gamma}{\frac{{\cal A}(s,0^-)s^3 ds}{(s^2+\mu^4)^3}}
\end{equation}
where the value $0^-$ must be understood as $t\rightarrow 0^-$ at all times. Thus, we retain only divergent and finite terms, and get rid of those that vanish polynomially in $t$. Here the integral is taken over the contour $\gamma$, as shown in figure \ref{fig:gamma_cont}, corresponding to the sum of two small circles surrounding the values $s=\pm i \mu^2$. The choice of $\mu \in \mathbb{R}$ is a matter of convenience, and later we will assume it to be much smaller than the cutoff scale of the low energy EFT.

Note that the analytic structure of the amplitude is completely determined by the assumptions above. It is fully analytic in the whole $s$-plane, except for the branch cut along the whole real axis. We can thus modify the integration contour to $\Gamma$, consisting on two lines ${\rm Re}(s)\pm i\epsilon$, with $\epsilon\ll 1$, together with two arcs of infinite radius. Thus
\begin{equation}
\label{eq:final_disp}
   \Sigma(\mu ,0^-)= \frac{1}{2\pi i}\oint_{\Gamma}{\frac{{\cal A}(s,0^-)s^3 ds}{(s^2+\mu^4)^3}} =\int_0^{\infty}\frac{ds}{\pi}\left({\frac{s^3\, {\rm Im}{\cal A}(s+i\epsilon,0^-)}{(s^2+\mu^4)^3}}+\frac{(s-4m^2)^3{\rm Im}{\cal A}^\times(s+i\epsilon,0^-)}{((s-4m^2)^2+\mu^4)^{3}}\right)+\Sigma_{\infty}.
\end{equation}

Here we have used the Schwarz reflection principle $A^*(s)=A(s^*)$ to relate the integral in the lower part of the complex plane to that on the upper part, and introduced the crossing-symmetric process\footnote{We have also used the fact that $s+t+u=4m^2$. For a detailed derivation of this expression, cf. \cite{Herrero-Valea:2020wxz}.} ${\cal A}^{\times}(s,t)$, obtained by letting $s\rightarrow u$, together with a change of variables, to rewrite the whole expression as an integral over positive values of $s$.

In the previous formula, $\Sigma_{\infty}$ stands for the sum of the integrals along the two infinite arcs $\Gamma_C=\Gamma^+ +\Gamma^-$ in the upper and lower parts of the complex plane,
\begin{align}
\label{sigma_inf}
   \Sigma_{\infty}=\frac{1}{2\pi i}\oint_{\Gamma_C}ds \frac{{\cal A}(s,0^-)}{s^{3}},
\end{align}
where $\mu$ has been neglected, since for this integral $|s|\rightarrow \infty$. The contribution of $\Sigma_\infty$ is normally ignored by invoking the Froissart-Martin bound \cite{Martin:1965jj} in the case of exchange of massive particles. For graviton exchange, it is typically assumed that a certain version of this bound holds in the form
\begin{align}\label{eq:froisart_martin}
    \lim_{|s|\rightarrow \infty}\left|\frac{{\cal A}(s,0^-)}{s^2}\right|=0,
\end{align}
which seems enough for the arc integrals to vanish. This bound has been rigorously derived for theories in $d>4$ space-time dimensions, but the realistic case of $d=4$ remains elusive, so eq.\eqref{eq:froisart_martin} has to be regarded as an extra assumption at this stage. In this work, we do not want to make such beforehand assumption on the UV behavior of the scattering amplitude, and thus we keep the integral arbitrary hereinforward. Our only starting requirement will be that $\Sigma_{\infty}$ does not contain any forward limit singularities stronger than a pole. Namely $\left(t\cdot  \Sigma_{\infty}\right)_{t\rightarrow 0^-}\sim \text{constant}$.

Note that \eqref{eq:final_disp} can be thought as a formula connecting the IR and UV behaviors of a given theory. While the RHS is an explicit integral along the full range of $s$, and thus sensitive to the properties of the UV theory; the value of $\Sigma(\mu,0^-)$ can also be computed in the IR region by using \eqref{S}, provided that $\mu$ is sufficiently small. In this case, it can even be computed in an EFT approximation of the full theory, as long as $\mu\ll \Lambda$, with $\Lambda$ being the cut-of of the EFT. This leads to a simple expression in terms of the residues of the integrand in \eqref{S}
\begin{equation}
\label{sigma}
\Sigma(\mu,0^-)=\frac{{\cal A}_{ss}(i \mu^2, 0^-)}{16}-\frac{3 i {\cal A}_s(i \mu^2,0^-)}{16 \mu^2},
\end{equation}
where ${\cal A}_s(x,0^-)=\left.\partial_s {\cal A}(s,0^-)\right|_{s=x}$, and equivalently with ${\cal A}_{ss}(x,0^-)$ and the second derivative. For an amplitude of the form \eqref{amplitude}, which can be obtained from an EFT coupled to General Relativity\footnote{Or, in general, to any theory whose tree-level gravitational dynamics matches that of the Einstein-Hilbert action.}, one obtains
\begin{equation}
    \Sigma(\mu,0^-)=\frac{1}{2}\left(\frac{A_0}{t}+A_1 \log{t}\right)+(\text{regular terms})+(\text{higher loops}).
\end{equation}

Thus, $\Sigma(\mu,0^-)$ defined as in eq. \eqref{S} captures exactly the coefficient in front of the $s^2$ term in the amplitude. Remarkably, this expression still contains singularities in the forward scattering limit $t\rightarrow 0^-$, produced by the $s^2$ dependence in those terms coming from graviton exchange. In the next section we discuss what these singularities tell us about the UV theory.

It is important to remark here that this problem is particular to graviton scattering. If one scatters any other massless species -- scalars or vector bosons -- which are instead described by a renormalizable theory, the scattering amplitude will still be divergent in the forward limit. However, these divergences come without a quadratic $s$-dependence, which means that they disappear when computing $\Sigma_\infty$, leading to a regular dispersion relation.

\section{Graviton scattering and cancellation of divergences}
\label{sec:asympt_exp}

Let us start by recalling that \eqref{eq:final_disp} is \emph{exact}. On its derivation there is no approximation or expansion whatsoever, besides taking an approximate forward limit. This means that, if the LHS is divergent, the RHS must be so. However, the latter depends only on the imaginary part of the scattering amplitude, which is regular within the physical region by application of the optical theorem
\begin{align}\label{eq:optical_theorem}
    {\rm Im}{\cal A}(s,0)=s\sqrt{1-\frac{4m^2}{s}}\ \sigma(s),
\end{align}
and the requirement that the total cross-section is finite. As discussed in \cite{Alberte:2021dnj}, this is enough to conclude that the divergence in the RHS of \eqref{eq:final_disp} must come from the failure of the integral to converge when $t\rightarrow 0^-$ at some high-energy regime $s\gg M_*^2$, where $M_*$ is thus the scale above which the EFT fails to describe graviton scattering and must be replaced by its UV completion. Assuming the mildest possible analytic behavior of the amplitude leads to a linear Regge trajectory -- cf. \cite{Alberte:2021dnj} and the appendix in \cite{Herrero-Valea:2020wxz}
\begin{align}
  \left. {\rm Im} {\cal A}(s,t)\right|_{s\gg M_*^2}=r_* s^{2+\alpha' t}.
\end{align}
Here $\alpha'\sim M_*^{-2}$ and $r_*$ are provided by the concrete UV completion leading to this form. By assuming this \emph{exact} high energy behavior, we can cancel the tree-level pole in the LHS of \eqref{eq:final_disp}. Such a linear Regge behavior for graviton scattering is typical in String Theories. Indeed, it can be obtained from the famous Veneziano amplitude \cite{Veneziano:1968yb}, describing the scattering of four closed bosonic strings \cite{Gross:1987ar}.

However, it is naive to assume that the Regge behavior at high energies is an exact linear trajectory. Indeed, this would lead to two problems. First, it only provides cancellation of the tree-level pole. Moreover, plausible candidates for UV amplitudes, like the scattering of strings, are not Regge exact, only their leading behavior is of this form. It is then natural to wonder what is the possible allowed form of these sub-leading corrections. One first mandatory requirement is that they must be able to cancel the logarithmic divergences as well as the pole. As shown in \cite{Herrero-Valea:2020wxz}, this requires sub-leading terms to contain a piece
\begin{align}
     \left. {\rm Im} {\cal A}(s,t)\right|_{s\gg M_*^2}=r_* s^{2+\alpha' t}\left(1+\frac{\zeta}{\log(\alpha's)}\right),
\end{align}
with $\zeta$ a dimensionless constant, but nothing else is known beyond this. We show now however, that we can indeed obtain a good amount of extra information on the UV amplitude by simply requiring the cancellation of divergences, obtaining some results that go beyond the linear Regge trajectory.

In order to proceed, we assume that the high energy form of the imaginary part of the amplitude reads
\begin{align}
    \left. {\rm Im} {\cal A}(s,t)\right|_{s\gg M_*^2}=s^{2+\alpha' t} \phi(s,t),
\end{align}
where $\phi(s,t)$ is an arbitrary function. We now take \eqref{eq:final_disp} and split the integration regime in the RHS as
\begin{align}
    \int_{0}^{\infty}=\int_{0}^{M_*^2}+\int_{M_*^2}^{\infty}.
\end{align}

As we previously discussed, all the divergences in this RHS come from the high-energy behavior of the integral, which means that the first piece in the previous expresion is regular. Thus, we move it to the LHS and write
\begin{align}
    \Sigma-\Sigma_{\infty}-\frac{1}{\pi}\int_{0}^{M_*^2} ds\  F(s)=\frac{1}{\pi}\int_{M_*^2}^\infty ds\ F(s),
\end{align}
where we have introduced
\begin{align}
    F(s)=\frac{s^3{\rm Im}{\cal A}(s+i\epsilon,0^-)}{(s^2+\mu^4)^{3}}+\frac{(s-4m^2)^3{\rm Im}{\cal A}^\times(s+i\epsilon,0^-)}{((s-4m^2)^2+\mu^4)^{3}}.
\end{align}

The LHS here thus contains divergences and regular pieces, that we decide to separate as
\begin{align}
     \lim_{t\rightarrow 0^-}
     \left(\Sigma-\Sigma_{\infty}-\frac{1}{\pi}\int_{0}^{M_*^2} ds F(s)\right)= \frac{\beta_0}{t}+\beta_1 \log(-t) + \bar{f},
\end{align}
where $\bar{f}$ is constant. On the other hand, we take the limit $\{m,\mu\} \ll M_*$ in the RHS, as well as the assumption that the external states are bosonic -- and thus ${\cal A}^{\times}(s,0^-)={\cal A}(s,0^-)$ -- arriving to
\begin{align}
 \frac{\beta_0}{t}+\beta_1 \log(-t) + \bar{f}=\frac{1}{\pi}\int_{M_*^2}^\infty \frac{ds}{s}s^{\alpha't} \phi(s,t)=\frac{M_*^{2\alpha't}}{\alpha' \pi}\int_0^\infty  d\sigma\ \phi(\sigma ,t) e^{\sigma t},
\end{align}
where in the last step we have performed a change of variables $s=M_*^2\  e^{\sigma/\alpha'}$. Finally, taking $x=-t$, reminding that the physical region corresponds to $t<0$, and thus $x>0$, and absorbing proportionality coefficients onto the definition of $\beta_0$, $\beta_1$ and $\bar f$, we arrive to the final expression
\begin{align}\label{eq:laplace_formula}
    \frac{\beta_0}{x}+\beta_1 \log(x) +\bar{f}=\int_0^\infty d\sigma\  \phi(\sigma,x) e^{-\sigma x},
\end{align}
where we can recognize the Laplace measure in the integral on the RHS.

Knowing the UV behavior of ${\rm Im}{\cal A}(s,t)$, then \eqref{eq:laplace_formula} can easily be used to compute the coefficients $\beta_0$, $\beta_1$ and $\bar f$, in a standard way. However, the inverse problem, obtaining the form of $\phi(\sigma,x)$ from the coefficients of the IR amplitude, is not so simple. Actually, this mathematical problem has, in general, infinitely many possible solutions, but not all of them will satisfy the analiticity properties that we must require for a physical amplitude\footnote{Indeed, a trivial solution is given be $\phi(\sigma,x)=e^{\sigma x}\left(\frac{\beta_0}{x}+\beta_1 \log(x) +\bar{f}\right)\delta(\sigma-x)$, which does not satisfy analiticity at $x=0$ for all values of $\sigma$.}.

In order to find a proper solution, let us make use here of Watson's lemma \cite{https://doi.org/10.1112/plms/s2-17.1.116} for the integral in \eqref{eq:laplace_formula}. We will thus assume that the function $\phi(\sigma,x)$ satisfies\footnote{Exponential boundedness is a softer behavior than the one required by the Froissart-Martin bound. The latter is actually problematic when confronted to the full Veneziano amplitude for string scattering, which does not satisfy it. Instead, Veneziano's amplitude is exponentially bounded. Exactly as we require here.}
\begin{align}
    \lim_{\sigma\rightarrow \infty}\frac{\phi(\sigma,x)}{e^{\gamma \sigma}}=0,
\end{align}
for some $\gamma>0$, and that it is a meromorphic function\footnote{The assumption of meromorphicity is true at one-loop, as we show below. However, we might be forced to abandon it in order to account for higher loop divergences in the IR, such as $\log(\log(x))$. Nevertheless, all these terms will enter with an extra scale suppression and we thus ignore them hereinafter.} around $\sigma=0$. Thus, it can be expanded in a Laurent series around this point
\begin{align}
    \phi(\sigma,x)=\sum_{n=0}^{\infty}\left(a_n(x)\sigma^n+\frac{b_n(x)}{\sigma^n}\right),
\end{align}
with $b_0(x)=0$. Any individual term of this sum, when plugged onto \eqref{eq:laplace_formula}, corresponds to a Laplace transform of $\sigma^a$ for a certain power $a$. 

Let us focus first on the non-analytic pieces of the series. By performing the integral -- using the analytic continuation of the $\Gamma$-function -- we get
\begin{align}
   b_n(x) \int_\epsilon^\infty d\sigma \ e^{-\sigma x} \sigma^{-n}= \frac{(-1)^n}{(n-1)!}\  b_n(x) x^{n-1}\log(x)+{\cal O}\left(x,\epsilon^{-1}\right),
\end{align}
where $\epsilon\ll 1$ is a regulator. For $n\geq 2$ we get terms which are not present in the LHS of \eqref{eq:laplace_formula}, unless $b_n(x)\sim x^{1-n}$. However, this violates the assumption of analiticity of $\phi(\sigma,x)$ when $x\rightarrow 0$. We conclude that all singular terms in the Laurent series for $n\geq 2$ must vanish. We thus have
\begin{align}\label{eq:Laurent}
    \phi(\sigma,x)=\frac{b_1(x)}{\sigma}+\sum_{n=0}^{\infty}a_n(x)\sigma^n,
\end{align}
where $b_1(x)=\beta_1+{\cal O}(x)$, in order to account for the $\log(x)$ forward divergence in the LHS of \eqref{eq:laplace_formula}. This justifies the choice done in \cite{Herrero-Valea:2020wxz}.

We shift our focus now to the Taylor series. Since $\phi(\sigma,x)$ is analytic around $x=0$, we have
\begin{align}\label{eq:cond_cancel}
    \lim_{x\rightarrow 0} a_n(x)=a_n x^{\eta_n},
\end{align}
where all the $\eta_n$ are constant and we assume them to be different. Later we will see that this is necessary to avoid double and higher poles in \eqref{eq:laplace_formula}. For now, let us take it as an assumption. We invoke now the partial wave expansion of the amplitude for a unitary theory, which implies -- see Appendix B of \cite{deRham:2017imi} for a proof
\begin{align}
     \frac{d^k}{dt^k}\left. {\rm Im}{\cal A}(s,t)\right|_{t=0}\geq 0,
\end{align}
for all $k$ and all values of $s$ within the physical region. Using this fact, we easily conclude by direct computation that 
\begin{align}
    a_n\geq 0, \quad \text{for all}\ n.
\end{align}

Knowing this, we plug now the Taylor series in \eqref{eq:Laurent} back onto the integral, and by integrating term by term we get
\begin{align}\label{eq:integral_sum}
 \sum_{n=0}^{\infty} \int_0^\infty ds\ e^{-s x}a_n(x)s^n=\sum_{n=0}^\infty \frac{a_n(x) \Gamma(n)}{x^{n+1}}.
\end{align}

Since all $a_n>0$, there cannot be cancellations between different values of $n$, which implies that all the terms in the RHS must, at most, diverge as a single pole. This implies
\begin{align}\label{eq:cond_an}
    a_n(x)=a_n x^n+{\cal O}\left(x^{n+1}\right),
\end{align}
for some constant, perhaps vanishing, coefficient $a_n$. Note however that, in order to cancel the single pole $\beta_0/t$ in \eqref{eq:laplace_formula}, at least one of the $a_n$ coefficients must be non-vanishing.

At this point one might be worried about the convergence of the sum in \eqref{eq:integral_sum}, since we are expanding around $\sigma=0$ and integrating over the whole real line. However, this leads to no problems in the setting discussed here. Let us show this explictly by cutting the Taylor series at a finite order $N$
\begin{align}
    \phi(\sigma,x)=\frac{b_1(x)}{\sigma}+\sum_{n=0}^N a_n(x)\sigma^n+\mathfrak{R}_{N+1}(\sigma,x).
\end{align}
Since this is a Laurent series, there exist a function $K(x)$ such that
\begin{align}
    |\mathfrak{R}_{N+1}(\sigma,x)|< K(x) \sigma^{N+1},
\end{align}
at least in the limit $x\rightarrow 0$ of interest. Using this condition we can thus estimate the size of the remainder after cutting the series and exchanging the order of summation and integration. We have
\begin{align}
  \left|\int_0^\infty d\sigma\ e^{-\sigma x}\mathfrak{R}_{N+1}(\sigma,x) \right|< \int_0^\infty d\sigma\ e^{-\sigma x}\left|\mathfrak{R}_{N+1}(\sigma,x)\right| <K(x) \int_0^\infty d\sigma\ e^{-\sigma x}\sigma^{N+1}.
\end{align}
The last integral is immediate and gives
\begin{align}
     \left|\int_0^\infty d\sigma\ e^{-\sigma x}\mathfrak{R}_{N+1}(\sigma,x) \right|<{\cal O}\left(\frac{1}{x^{N+2}}\right).
\end{align}
Noting that the $N$-th term in the series contributes at order $a_n(x)x^{-(N+1)}$, this shows that \eqref{eq:integral_sum} is thus well-behaved as an asymptotic series, which is enough for our purposes here.

Before going forward let us go back to the condition \eqref{eq:cond_cancel}. Now it is obvious that all $\eta_n$ have to be different in the limit $x\rightarrow 0$. Unless they satisfy \eqref{eq:cond_an} there would be extra divergences after integration of $\phi(\sigma,x)$. A possible way out would be to have two terms giving rise to the same divergence and cancelling each other. However, since all $a_n>0$, this is not possible. We thus conclude that the form of our asymptotic expansion is indeed unique and reads in the forward limit, once all knowledge is collected
\begin{align}\label{eq:phi_forward_limit}
   \lim_{x\rightarrow 0} \phi(\sigma,x)=\frac{\beta_1}{\sigma}+\sum_{n=0}^\infty a_n (x\sigma)^n, \quad \sum_{n=0}^\infty a_n \Gamma(n)=\beta_0,
\end{align}

Note that the expansion of the function $\phi(\sigma,x)$ in the forward limit, which naively corresponds to $x\rightarrow 0$, has become instead an expansion when $x\sigma\rightarrow 0$, since it is only under this assumption that \eqref{eq:phi_forward_limit} is well-behaved as an asymptotic expansion. This suggest that the proper forward limit to take in the presence of graviton exchange, at least at high energies, is actually
\begin{align}\label{eq:limit}
    \tau=\sigma x \sim t\log(s)\rightarrow 0,
\end{align} or $t\log|s|\rightarrow 0$ in the complex plane for $s$, which ensures perturbative control of the UV amplitude. As we will see in a moment, this has a strong impact on the derivation of positivity bounds. Since the asymptotic limit $|s|\rightarrow \infty$ and $t\rightarrow 0$ has to satisfy \eqref{eq:limit}, the computation of the integral along $\Gamma_C$ in \eqref{sigma_inf} has to be taken carefully.


\section{Arc integrals in the forward limit}
\label{sec:arc_int}
By means of analyticity and crossing symmetry, one can actually go beyond our results in the previous section and recover the asymptotics of the whole amplitude out of its imaginary part. Indeed, for $|s|\gg M_*^2$ we can write a totally standard dispersion relation for the scattering amplitude by using Cauchy's integral theorem. We have \cite{deRham:2017avq}\cite{Herrero-Valea:2020wxz}
\begin{equation}
    {\cal A}(s,t)=\frac{s^2}{2\pi i}\oint_{\gamma_s}\frac{{\cal A}(z,t)dz}{z^2(z-s)}=F(s,t)+F(-s-t,t).
\end{equation}
Here $\gamma_s$ is a small circle around $z=s$ and we have exploited crossing symmetry to obtain an explicit expression in terms of $s$ and $u=-s-t$. The function $F(s,t)$ reads\footnote{See \cite{deRham:2017avq} for a detailed derivation. Here we are taking the subtraction point $\mu_p^2\ll |s|$, and taken into account that all pathologies of the scattering amplitude, such as the pole, are contained in the IR.}
\begin{equation}
    F(s,t)=\frac{ s^2}{\pi}\int_{0}^{\infty}{\frac{{\rm Im}{\cal A}(z,t)dz}{z^2(z-s)}}=\frac{ s^2}{\pi}\int_{0}^{M_*^2}{\frac{{\rm Im}{\cal A}(z,t)dz}{z^2(z-s)}}+\frac{ s^2}{\pi}\int_{M_*^2}^{\infty}{\frac{a_0 z^{\alpha' t}dz}{z-s}} +{\cal O}\left(t \log(s)\right),
\end{equation}
where we have taken into account only the leading term in the expansion \eqref{eq:phi_forward_limit}.

In the region of validity $|s|\gg M_*^2$, and the first integral becomes proportional to $s/M_*^2$, so that it can be neglected. The asymptotics of the last one for large $|s|$ leads instead to
\begin{equation}
    F(s,t)=-\frac{a_0 e^{-i\pi \alpha' t}}{\sin{(\pi \alpha't)}}s^{2+\alpha't}.
\end{equation}

Thus, we see that the total leading part of the amplitude, and not only its imaginary part, is actually completely determined. It reads
\begin{equation}
\label{eq:Regge}
    A(s,t)=-\frac{a_0 e^{-i\pi \alpha' t}}{\sin{(\pi \alpha't)}}(s^{2+\alpha't}+(-s-t)^{2+\alpha't})+{\cal O}\left(t\log (s)\right).
\end{equation}
This result of course reproduces the imaginary part $a_0 s^{2+\alpha' t}$, while being at the same time its analytic and crossing symmetric continuation\footnote{Notice that for large $s>0$ and small $t<0$ the second term does not contribute to the discontinuity.}.

After obtaining this asymptotic form for the UV scattering amplitude including exchange of gravitons, we can now focus on understanding whether we can really set the contribution of the infinite arcs $\Sigma_{\infty}$ to the dispersion relation \eqref{eq:final_disp} to vanish or not. In the case of gapped theories, the Froisart-Martin bound ${\cal A}(s,t)<s\log^2{s}$ guarantees that $\Sigma_{\infty}=0$. For \eqref{eq:Regge} instead, we find a different result.

Let us then compute the integral in \eqref{sigma_inf} explicitly. Taking into account that the arc $\Gamma_R$ is described by $s=R e^{i\theta}$ with $R\rightarrow \infty$, and that
\begin{align}
    \frac{1}{2\pi i}\oint_{\Gamma_R}ds \frac{s^{2+\alpha' t}}{s^{3}}=\frac{ R^{\alpha't}}{2\pi} \int_0^{2\pi} d\theta  e^{i\alpha't \theta}=\frac{ R^{\alpha't }}{2\pi}\frac{e^{2\pi i\alpha't}-1}{i\alpha't },
\end{align}
we get
\begin{equation}
    \Sigma_{\infty}= -\frac{2 a_0 e^{-i\pi \alpha' t}}{\sin{(\pi \alpha't)}} \frac{ R^{\alpha't }}{2\pi}\frac{e^{2\pi i\alpha't}-1}{i\alpha't }=-\frac{2 a_0}{\pi\alpha' t}R^{\alpha' t}=-\frac{2 a_0}{\pi\alpha' t},
\end{equation}
where we have taken into account that the asymptotic expansion is well-controlled only when $t\log R\rightarrow 0$. Since in this limit $R^{\alpha' t}\rightarrow 1$, we get a non-vanishing contribution, unlike in a case where one takes first the limit $R\rightarrow \infty$ with small but finite $t$ \cite{inpreparation}. 

For the sake of completeness, let us notice that if one takes the integral over the branch cut in \eqref{eq:final_disp} not from $M_*^2$ to infinity, but from $M_*^2$ to $R$, one obtains instead
\begin{equation}
    \Sigma_{UV}=\frac{2}{\pi}\int_{M_*^2}^{\infty}\frac{ds\,{\rm Im}{\cal A}(s,t)}{s^3} =\frac{2}{\pi}\int_{M_*^2}^{R}\frac{ds}{s} \,a_0 s^{\alpha' t}=\frac{2a_0}{\pi \alpha' t}\left(R^{\alpha' t}-(M_*^2)^{\alpha' t}\right).
\end{equation}

In the limit $t \log{R}\rightarrow 0$ both terms go to unity and we get
\begin{equation}
    \Sigma_{UV}=O(t)+O(M_*^4)
\end{equation}
without the $1/t$ divergence, which appears instead to be captured in the arc contribution $\Sigma_R$. If instead we carefully take $R\rightarrow \infty$ first at finite negative $t$, the familiar result with $\Sigma_{\infty}=0$ is reproduced \cite{inpreparation}. Thus, it seems that the combination $\Sigma_{UV}+\Sigma_{\infty}$ does not depend on the way of taking the limits in $R$ and $t$, although individual terms do. Hereinafter we use the limit $\tau\rightarrow 0$ motivated by our findings in section \ref{sec:asympt_exp}. This simplifies the computation by allowing us to use only the leading term in the $\tau$ expansion for the imaginary part of the amplitude. 

With this choice, and for the simple amplitude \eqref{eq:Regge}, the infinite arc brings a $1/t$ term. However, sub-leading terms might lead also to finite contributions. In particular, by noting that the new pole contribution comes from the real part of the amplitude, which is not constrained by unitarity or any other of the arguments here, it seems that any finite value can be obtained from the UV, simply by modifying ${\rm Re}{\cal A}(s,t)$. For example, take the equally valid amplitude
\begin{equation}
\label{eq:A_UV_simple}
A(s,t)=-\frac{a_0 e^{-i\pi \alpha' t}}{\sin{(\pi \alpha't)}}(s^{2+\alpha't}+(-s-t)^{2+\alpha't})+\beta e^{-i\pi \alpha' t}(s^{2+\alpha't}+(-s-t)^{2+\alpha't}),
\end{equation}
which leads to
\begin{equation}
  \Sigma_{\infty}=-\frac{2 a_0}{\pi\alpha' t} +\beta +O(t).
\end{equation}
This indeed contains a finite piece that must be matched by the infrared terms in the LHS of \eqref{eq:final_disp} in order for the dispersion relation to be valid. Notice however, that $\beta$ is not constrained at all, and in particular it is not forced to be positive. It can be large ($|\beta|\gg M_*^{-4}$) and negative. Thus, its value will influence applicability of positivity bounds based on the relation \eqref{eq:final_disp}. We will discuss this point later.

Here we did not include the $\log\left(-t\right)$ divergences in the dispersion relation. Given that they can be cancelled by the sub-leading contributions to ${\rm Im}{\cal A}(s,t)$ at large $s$, their impact to $\Sigma_R$ should be also sub-leading. An accurate computation shows that the term $s^{2+\alpha' t}/(\log{s})$ appearing in the imaginary part would always lead to a vanishing contribution. Thus, only those terms needed to cancel the leading IR divergences induce a non-trivial value of the infinite arc integrals. However, let us also note that although sub-leading IR divergences are not sensitive to the real part of the amplitude in the UV, they are still relevant to recover its imaginary part, as previously discussed in this work.


\section{QED with gravity}\label{sec:QED}
Following the derivation of positivity bounds from twice-subtracted dispersion relations, several works examined their consequences on different physical theories of interest for model building. In particular, bounds in the presence of graviton exchange were closely studied in recent works \cite{Alberte:2020bdz,Alberte:2021dnj}. There, and by looking at photon scattering, the authors show that positivity bounds are easily violated by gravitational contributions. In this section we examine how our findings and, in particular, the contribution of the infinite arcs, relax this issue.

\begin{figure}
    \centering
    \includegraphics[width=0.5\textwidth]{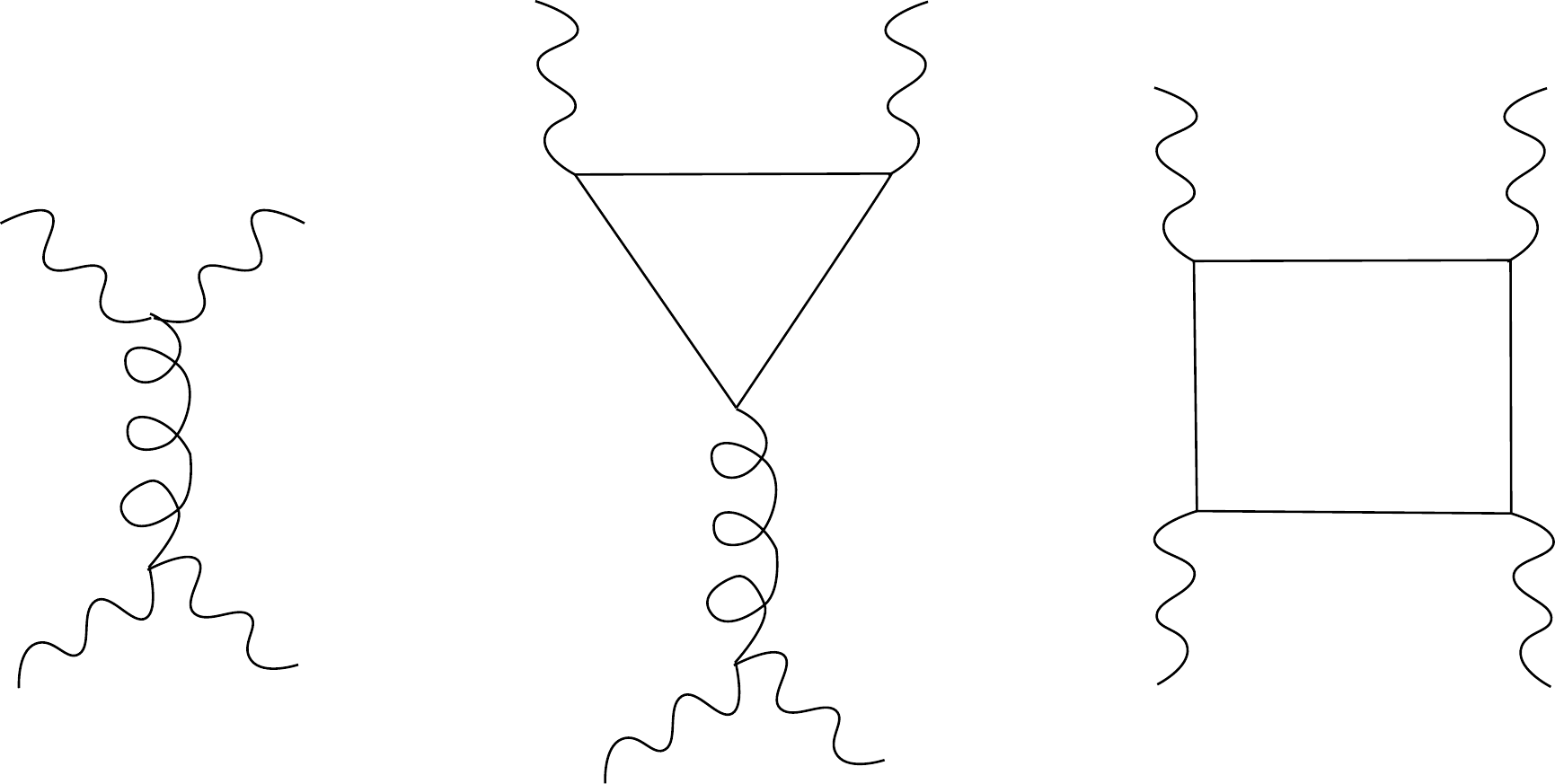}
    \caption{Topologies providing the leading contributions to the photon scattering amplitude at large $s$ in the forward limit.}
    \label{fig:diagrams_QED}
\end{figure}

Let us then consider Quantum Electrodynamics (QED) coupled to gravitation, with action
\begin{equation}\label{eq:QED}
    S=\int\sqrt{-g}\,d^4x\left(-\frac{M_P^2}{2}R-\frac{1}{4}F_{\mu\nu}^2+\bar{\psi}(D_{\mu}\gamma^{\mu}-m)\psi\right),
\end{equation}
where $F_{\mu\nu}$ is the photon field strength, and $\psi$ is a fermion field with charge $e$.  Following \cite{Alberte:2020bdz} we look at $2\rightarrow 2$ photon scattering, including tree-level graviton exchange and one-loop fermion corrections -- thus retaining contributions up to ${\cal O}\left(e^6\right)$ and ${\cal O}\left(M_P^{-4}\right)$. In the forward limit, and by taking $s\gg m^2$, this reads
\begin{equation}
    A(s,t)=-\frac{s^2}{M_P^2 t}+\frac{1}{M_P^2}\left(-\frac{11 e^2 s^2}{360 \pi^2 m^2}+\frac{e^2 s}{12\pi^2}\right)+\frac{11 e^4 s^2}{720 \pi^2 m^4}.
\end{equation}
The different topologies contributing to this scattering are shown in figure \ref{fig:diagrams_QED}.

The RHS of \eqref{eq:final_disp} contains four different pieces when computed for this amplitude. The first one is the integral of the imaginary part of the amplitude for those contributions that survive in the limit $M_P\rightarrow \infty$. These can be computed analytically, since the action \eqref{eq:QED} corresponds to a renormalizable theory in this limit, and were obtained in \cite{Alberte:2020bdz}. We will borrow their result here. The second piece is the contribution given by pure gravitational terms in the IR. Again, these can be computed explicitly by using the amplitude above, as \cite{Alberte:2020bdz} did. We also have those coming from the UV part of the integral, when $s\gg M_*^2$, and which depend on the UV completion, as previously discussed. We name them $\Sigma_{\rm UV}$. Finally, we have the contribution of the infinite arcs, which we have learnt that cannot be taken to vanish a priori. The total result for the right hand side -- in the limit of small $\mu$ -- is then
\begin{equation}
    \Sigma(\mu^2,0^-)=\frac{11 e^4 }{360 \pi^2 m^4}-\frac{11 e^2}{360 \pi^2 m^2 M_P^2}+\Sigma_{\rm UV}+\Sigma_{\infty}.
\end{equation}

On the other hand, the LHS of \eqref{eq:final_disp} can be directly computed and reads in this case
\begin{equation}
    \Sigma(\mu^2,0^-)=A_{ss}(s)=-\frac{2}{M_P^2 t}-\frac{11 e^2 }{180 \pi^2 m^2 M_P^2}+\frac{11 e^4}{360 \pi^2 m^4}.
    \end{equation}
    
As we can see, both results agree in the decoupling limit of gravitational interactions. This is not a surprise because, as we have already pointed out, the action \eqref{eq:QED} is renormalizable in this limit. This means that \eqref{eq:final_disp} becomes trivial. The divergence $1/t$ however, has a pure gravitational origin and, as we have discussed, will be cancelled by the interplay between $\Sigma_{UV}$ and $\Sigma_{\infty}$. However, the contributions of order $(m M_P)^{-2}$, showing up in both sides, do not cancel each other. If we were being naive, assuming that $\Sigma_\infty=0$ and simply cancelling the pole with $\Sigma_{\rm UV}$, then we would find a clash between the two approaches to derive $\Sigma(\mu^2,0^-)$. The only way out is to assume the existence of new physics turning on before gravitational interactions, such that the amplitude gets modified and leads to a cancellation of the undesired piece
\begin{align}
    -\frac{11 e^2 }{360 \pi^2 m^2 M_P^2}+\frac{\Theta}{\Lambda^2}=0,
\end{align}
where $\Theta$ is a constant. From simple dimensional analysis we can see that this implies that the cut-off of the SM -- in other words, the scale of introduction of new physics -- would be $\Lambda\sim \sqrt{m M_P}$, which in QED leads to $\Lambda \sim 10^8 {\rm GeV}$, significantly lower than the Planck scale. This option was studied in \cite{Alberte:2021dnj}.

However, there is another natural solution to the problem -- the possibility of having a non-zero infinite arc contribution $\Sigma_{\infty}$ with negative sign. As we have shown in previous sections, the value of $\Sigma_\infty$ contains contributions from the real part of the gravitational amplitude at high energies -- the term $\beta$ in \eqref{eq:A_UV_simple} --, which are not constrained at all. Thus, they can potentially cancel the remaining negative contribution of order $(m M_P)^{-2}$. Moreover, by considering this possibility, we can obtain non-trivial information about the UV completion of gravitational interactions. In particular, in the case of QED, we can conclude that the UV amplitude in the Regge limit should `know' about the presence of the light particle (electron), since it needs to contain a large negative contribution related to it. For instance, by borrowing expression \eqref{eq:A_UV_simple}, a possible consistent UV amplitude in the forward limit is
\begin{equation}
\label{A_QED}
A(s,t)=e^{-i\pi \alpha' t}\left(-\frac{11 e^2 }{360 \pi^2 m^2 M_P^2}-\frac{\pi \alpha' }{M_P^2\,\sin{(\pi \alpha't)}}\right)\left(s^{2+\alpha't}+(-s-t)^{2+\alpha't}\right),
\end{equation}
but this option is, of course, not unique and other possible amplitudes could lead to similar physical results -- cf. \cite{Alberte:2020jsk}. Let us stress that the Regge slope $\alpha'$ cannot be fixed from IR considerations, and it is instead connected to properties of the UV completion.

By considering this solution to the conundrum unveiled in \cite{Alberte:2020bdz,Alberte:2021dnj}, QED can be `rescued' and trusted as a good EFT when coupled to gravitation up to the Planck scale. Of course, a realistic model of QED breaks down before $M_P$, since it needs to be embedded onto the Electroweak model, but a similar reasoning can be made even in the general case of the Standard Model (SM) -- where the problem is even worse due to neutrino loops, which bring the cut-off down to values just slightly over the LHC reach. Finally, let us point out that the QED contribution to \eqref{A_QED} can be interpreted as the contribution of loops of light particles -- the fermion in this case -- to the Regge behaviour of the amplitude in the UV \cite{Alberte:2021dnj}.

\section{The fate of gravitational positivity bounds}
\label{sec:fate_bounds}
The prototypical application of formula \eqref{eq:final_disp} is to derive positivity bounds, constraints on the values of Wilson coefficients of EFTs, by computing explicitly the value of $\Sigma(\mu,0^-)$ in the IR. These are obtained by simply considering the following expression
\begin{equation}
    \Sigma=\int_{s_{\rm th}}^{\infty}\frac{ds}{\pi}\left({\frac{s^3\, {\rm Im}{\cal A}(s,0)}{(s^2+\mu^4)^3}}+\frac{(s-4m^2)^3{\rm Im}{\cal A}^\times(s,0)}{((s-4m^2)^2+\mu^4)^{3}}\right),
\end{equation}
where we have assumed that $\Sigma_\infty =0$ for the moment. Here $s_{\rm th}$ stands for the threshold of particle production where the branch cut starts on the real axis. For scattering processes without masless particles in the exchange channel, this corresponds to $s_{\rm th}=4m_l^2$, with $m_l$ the mass of the lightest exchanged state, and the integral runs along the physical regime for the Mandelstam variable $s$ \cite{Adams:2006sv}. In the case of a massless exchange, we have $s_{\rm th}=0$. Taking into account that the optical theorem \eqref{eq:optical_theorem} implies that the integrand in the RHS is always positive, from unitarity requirements of the UV completion, we can conclude that
\begin{align}\label{eq:positivity_b_simple}
    \Sigma>0,
\end{align}
which in turn will imply conditions on the Wilson coefficients contributing to the scattering amplitudes and ultimately to $\Sigma$.

The bounds \eqref{eq:positivity_b_simple} can be improved by noting that part of the RHS can actually be computed within an EFT. Splitting the integral in the RHS as $\int_{s_{\rm th}}^\infty=\int_{s_{\rm th}}^{\Lambda^2}+\int_{\Lambda^2}^{\infty}$, we can move the first piece to the left and conclude in the same fashion that
\begin{equation}
    \Sigma-\int_{s_{\rm th}}^{\Lambda^2}\frac{ds}{\pi}\left({\frac{s^3\, {\rm Im}{\cal A}(s,0)}{(s^2+\mu^4)^3}}+\frac{(s-4m^2)^3{\rm Im}{\cal A}^\times(s,0)}{((s-4m^2)^2+\mu^4)^{3}}\right)>0.
\end{equation}
These \emph{improved} positivity bounds have been also referred in the literature as \emph{beyond} positivity bounds \cite{Bellazzini:2017fep}.

If $s_{th}\ne 0$ there is a simpler way to bound the coefficient in front of $s^2$ in the amplitude. One can equivalently derive a bound for 
\begin{equation}
    \bar\Sigma=\frac{1}{2\pi i}\oint\frac{{\cal A}(s)\,ds}{(s-\mu^2)^3}=\frac{1}{2}{\cal A}_{ss}(s)>0,
\end{equation}
which is applicable for $\mu^2<s_{th}$ \cite{Adams:2006sv,deRham:2017xox,Bellazzini:2017fep}. This approach, however, cannot be directly applied for scattering of massless particles. The presence of a branch cut with $s_{\rm th}=0$ requires to use the more complicated dispersion relation \eqref{eq:final_disp}.

These type of bounds can be systematically obtained from many different dispersion relations, by simply taking more subtractions -- see \cite{deRham:2017avq, Bellazzini:2020cot}. Even amplitudes containing graviton exchange can provide rigorous bounds for the coefficients in front of higher powers of $s$ -- $s^4$ and beyond --, as well as for their $t$ derivatives \cite{Bern:2021ppb}, which are regular in the forward limit. This happens because the $1/t$ pole (as well as the loop IR singularities) is accompanied by a $s^2$ power at most\footnote{This may not be true in theories with higher spin states, which we are not considering here.}. For this reason, only the application of positivity bounds for the $s^2$ term gets obstructed in the presence of graviton exchange and graviton loops.

Of course, one can proceed naively by regularizing the divergence in the same way as we have proceeded here, by keeping small $t<0$, and simply bound the coefficient accompanying the divergence to be negative, since it dominates the bound. However, this is just the residue in the pole of the graviton propagator, whose sign is already constrained to satisfy trivial requirements of perturbative unitarity. Thus, no new information is obtained from positivity bounds in this case, unless one resolves the singularity. This can be done provided that the contribution of the infinite arc $\Sigma_\infty$ vanishes. In this case the divergence can be simply cancelled by the proper Regge behaviour in the UV. However, there are finite remainders whose sign cannot be determined a priori, and thus one arrives to an approximate positivity bound
\begin{equation}
    \Sigma>-O(M_*^{-4}).
\end{equation}
which allows for small negativity \cite{Tokuda:2020mlf,Herrero-Valea:2020wxz}.

As we have mentioned, this is only true under the extra assumption that the infinite arc contribution is either zero in the limit $t\rightarrow 0$ or shown to be parametrically smaller than $O(M_*^{-4})$. However, as we have discussed in section \ref{sec:QED} with the example of QED coupled to gravity, the contribution of the infinite arc can actually be negative and parametrically large, violating this assumption. This reflects the fact that the loops of light particles can affect the amplitude in the forward limit even in the UV region of large $s$. If this happens, then no bounds can be set for the finite part of the $s^2$ term in the amplitude, since the former are modified to
\begin{align*}
    \Sigma>\Sigma_\infty -{\cal O}\left(M_*^{-4}\right),
\end{align*}
which is meaningless without a systematic way to determine the size of $\Sigma_\infty$. Any amount of negativity can always be explained by contributions to the UV amplitude which, to the best of our knowledge, do not contradict any of the basic principles of QFT. 

Although the identification of the undetermined term as part of the infinite arc integral is related to our choice of kinematics in the UV, controlled by $\tau\rightarrow 0$, let us stress that the previous conclusion is not tied to it. For other choices, $\Sigma_\infty$ might vanish, but a similar contribution would arise from the branch cut, leading to the same physical conclusion \cite{inpreparation}.

As a final note, let us note that a possible way out to this conundrum is the case when the IR amplitudes are parametrically larger than the UV contribution to the arcs. This requires the existence of a cut-off scale in the IR $\Lambda$ such that the amplitude in this region can be organized as an EFT
\begin{align}
    {\cal A}(s,t)=\sum_{n=0}^\infty {\cal A}_n(s,t) \Lambda^{-n},
\end{align}
while gravitational dynamics will contribute with terms ordered by of $M_P$. In the case in which $\Lambda \ll M_P$, the contribution of $\Sigma_\infty$ can thus be safely neglected, so that we recover an approximate positivity bound
\begin{align}
\Sigma>{\cal O}\left(M_P^{-2}\right).
\end{align}

This justifies the application of positivity bounds to the case of gapped theories much below the gravitational scale -- and the neglection of gravity even though everything universally couples to gravitation --, but the problem survives if one wants to account for graviton exchange. Information about the UV completion is needed.

\section{Do string theories provide bona-fide positivity bounds?}\label{sec:strings}

We cannot say that we are currently in position to provide a number of known non-perturbative amplitudes for the test of our findings above for the UV behavior of gravitational scattering. However, string theory gives us some hints on how some examples are constructed.
In this section we use string amplitudes as a test ground to see if one indeed recognizes $\tau=t\log(s)$ as the expansion parameter of the forward limit, and to assess what happens to the large arc integral, i.e. if they can provide a constant contribution, similar to the one that solves the QED conundrum in section \ref{sec:QED}, upon some conditions or not.

Let us give a try to the 4-graviton scattering amplitude derived in type II superstring theory. 
It can be written in the following form \cite{Schwarz:1982jn,Tokuda:2020mlf}
\begin{equation}
    {\cal A}_{\rm string}(s,t)=-A(s^2t^2+s^2u^2+t^2u^2)\left.\frac{\Gamma(-s)\Gamma(-t)\Gamma(-u)}{\Gamma(1+s)\Gamma(1+t)\Gamma(1+u)}\right|_{u=-s-t}
    \label{ssa}
\end{equation}
where $A$ is positive and the string constant is set to $\alpha'=4$ for simplicity. This amplitude represents the tree-level 4-graviton scattering in the closed super-string NS-NS sector. The polynomial factor accounts for the polarisation states of the spin-2 particle. This result has some limitations, though. Neither R-NS nor open-closed string interactions are included here. Moreover, it takes into account only scattering of string states without touching the question on how branes enter into the game. Note that D-branes, which superficially accommodate the previously successful Chan-Paton factors, are responsible for the inclusion of SM particles in string phenomenology scenarios \cite{Polchinski:1996fm,Douglas:1996uz}. All this means that the amplitude above is far from describing our real world, but it is still of great interest for our purposes here, because it is an example of an extremely successful theoretically justified non-perturbative scattering description.

The expansion of the amplitude (\ref{ssa}) for $s\to\infty$ and $t\to0$ can be performed straightforwardly. Arranging the terms in powers of $t$ and keeping the leading $s$ contribution we get
\begin{equation}
    {\cal A}_{\rm string}(s,t)=A\frac{s^2}t\left(1+2t\log(s)+2t^2\log^2(s)+\frac43t^3\log^3(s)+\frac23t^4\log^4(s)+\dots\right),
\end{equation}
where we readily reveal the canonical $s^2/t$ pole for the spin-2 massless particle, and recognize the presence of $\tau=t\log(s)$ as the expansion parameter. We note that the appearance of $\tau$ is a very non-trivial property. It was not granted a priori to observe it here. However, our considerations in previous sections suggested its presence as a necessary condition for a healthy amplitude. Its emergence here thus serves as a very non-trivial sanity check of our results.

Computing $\Sigma(\mu,0^-)$ out of ${\cal A}_{\rm string}(s,t)$ one gets
\begin{equation}
    \Sigma(\mu,0^-)=-\frac A{2t}+4At+O\left(t^2\right),
\end{equation}
where we notice that the first sub-leading contribution is linear in $t$, with no constant ${\cal O}(1)$ term. This result can be obtained by following the procedure outlined in \cite{Tokuda:2020mlf}. It requires summing over residues of poles arising at the integer negative points of the $\Gamma$-function. Remarkably, note that since the constant term is absent, the amplitude \eqref{ssa} does not provide the large negative contribution that saves the day in the case of QED -- cf. section \ref{sec:QED}.

Therefore, we conclude that pure NS-NS superstring amplitudes cannot heal the curious contribution observed in \cite{Alberte:2020bdz}. However, there is hope for this to happen once SM particles are included in the amplitude, through coupling to D-branes. This is definitely an ambitious open question to be understood in the string framework, as one needs generalizations of SM amplitudes computed from the string perspective. Alternatively, there may still be a window in the string framework on its own if one includes other contributions arising from NS-R interactions or from open-closed string interactions. This analysis is however  clearly beyond the scope of the present work.

\section{Conclusions}
\label{sec:conclusions}
In this paper we have shown that the requirement of cancellation of IR forward divervenges appearing in graviton (mediated) scattering is enough to constrain the form of the imaginary part of the scattering amplitude at very high energies, above a scale $M_*$. In particular, we have proven that whatever the form of the UV completion of gravitation is, ${\rm Im}{\cal A}(s,t)$ must admit an asymptotic expansion of the form \eqref{eq:phi_forward_limit} in the limit $\tau\propto t \log s\rightarrow 0$. The appearance of the parameter $\tau$ is a highly non-trivial feature that is however reproduced in the known case of the Veneziano amplitude of string theory \cite{Veneziano:1968yb}, as we discuss in section \ref{sec:strings}. 

The determination of the form of ${\rm Im}{\cal A}(s,t)$ has an inmediate impact on the construction of positivity bounds, which are widely used to constrain EFTs of matter coupled to gravitation. In their derivation, there appear integrals along arcs with radius $|s|\rightarrow \infty$, which are typically taken to vanish, either by invoking the Froisart-Martin bound for gapped theories, or with other arguments in the gapless case. By using our expansion we show however that this cancellation is not guaranteed and instead depends on the form of the \emph{real} part of the scattering amplitude, which is not constrained at all, to the best of our knowledge. In the case that this real contribution exists, then the predictability of gravitational positivity bounds is doomed, since the previous simple expression $\Sigma>0$, which is computable within an EFT, gets modified into 
\begin{align}
    \Sigma-\Sigma_\infty>0,
\end{align}
which is meaningless unless some input about the contibution of the UV completion $\Sigma_\infty$ is given case by case.

Although this situation is overall negative for the applicability of positivity bounds, it can also have a bright side. The undetermined contribution from the UV completion could compensate the presence of anomalously large negative terms appearing in $\Sigma$ in the case of QED coupled to Einstein-Hilbert gravity, which we have studied in section \ref{sec:QED}, and in the general case of the SM. A naive solution to both preserving the fate of positivity bounds, and accepting these terms, is to assume the existence of new physics above a relatively low energy scale $\Lambda\sim \sqrt{m_l M_P}$, where $m_l$ is the mass of the lightest fermion. In the case of the SM this can be within the LHC scale and thus puts in tension the validity of the SM itself. In contrast to this solution, the existence of a non-vanishing contribution $\Sigma_\infty$, coming from the real part of the scattering amplitude of the UV completion, can solve the issue and preserve the validity of the SM up to the Planck scale. However, this would imply that positivity bounds cannot give us any new information in these situations.

Alternatively, we can look at this as an opportunity to realize a reverse bootstrapping. We can use the theories in the IR to compute the value of $\Sigma$, and use it to determine contributions to ${\rm Re}{\cal A}(s,t)$ at high energies through $\Sigma_\infty$. This could give an important insight on how light particles contribute to graviton scattering even beyond the Planck scale.

Finally, we have tested our results by looking at the non-perturbative amplitude for graviton scattering obtained from the scattering of NS-NS closed superstrings. This amplitude indeed organizes as an asymptotic expansion in $\tau\rightarrow 0$ in the double limit $s\rightarrow \infty$ and $t\rightarrow 0$, satisfying our results. However, it does not provide the negative large term $\Sigma_\infty$ required to save QED and the SM from a low cut-off. Although this could be interpreted as a hint of the true existence of this cut-off, we believe instead that it points out to the necessity of a better understanding of scattering amplitudes in the string framework. In particular, the problem at hand seems to require to go beyond the simple Veneziano amplitude and also account for interaction with SM particles through attaching the strings to D-branes, or perhaps including also NS-R sectors, and open-closed string interactions. Additionally, it is also interesting to question if there exist other UV completions besides string scattering that satisfy the requirements discussed in this work. Even more, we wonder if it is possible to construct model-independent amplitudes that not only cancel the IR forward divergences in graviton scattering, but also render the SM safe until the Planck scale, and what we can say about these amplitudes. 

One possible direction of research along these lines could be to consider triple-product amplitudes. Indeed, by looking at \eqref{ssa} we see that, apart from the polarization factor, the expression is a triple product of ${\cal B}(z)=\Gamma(-z)/\Gamma(1+z)$ such that ${\cal A}(s,t)\sim{\cal B}(s){\cal B}(t){\cal B}(u)$. Recently a new set of amplitudes with triple-product structure was considered in \cite{Huang:2022mdb}, where they claim that a wide class of functions ${\cal B}(z)$ leads to a unitary construction. It would be interesting to see whether those new amplitudes obey the constraints obtained in the present paper using a general model independent approach.


\section*{Acknowledgements}
We are grateful to Claudia de Rham and Andrew J. Tolley for discussions. M. H-V. is supported by the Spanish State Research Agency MCIN/AEI/10.13039/501100011033 and by the EU NextGenerationEU/PRTR funds, under grant IJC2020-045126-I. IFAE is partially funded by the CERCA program of the Generalitat de Catalunya. A. S. K. is supported in part by FCT Portugal investigator project IF/01607/2015. A. T. is supported by the European Union Horizon 2020 Research Council Grant No. 724659 MasssiveCosmo ERC2016COG and by a Simons Foundation Award ID 555326 under the Simons Foundation Origins of the Universe initiative, Cosmology Beyond Einstein’s Theory.

\bibliography{biblio}{}

\end{document}